\documentclass[%
 reprint,
groupedaddress,
showpacs,preprintnumbers,
amsmath, amssymb, aps,color
]{revtex4-1}

\usepackage{graphicx}
\usepackage{dcolumn}
\usepackage{bm}
\usepackage{color}

\newcommand{\tp}[1]{\textcolor{red}{#1}}

\begin{document}

\title{Bragg solitons in nonlinear $\mathcal{P}\mathcal{T}$-symmetric periodic potentials}

\author{Mohammad-Ali Miri$^{1,*}$, Alejandro B. Aceves$^{2}$, Tsampikos Kottos$^{3}$, Vassilios Kovanis$^{4}$, and Demetrios N. Christodoulides$^{1}$}
\affiliation{$^{1}$CREOL$/$College of Optics, University of Central Florida, Orlando, Florida 32816, USA\\$^{2}$Department of Mathematics, Southern Methodist University, Dallas, TX 75275, USA\\$^{3}$Department of Physics, Wesleyan University, Middletown, Connecticut 06459, USA\\$^{4}$Air Force Research Laboratory, Sensors Directorate, Wright Patterson AFB, OH 45433, USA}
\date{\today}
\begin{abstract}
It is shown that slow Bragg soliton solutions are possible in nonlinear complex parity-time ($\mathcal{PT}$) symmetric periodic structures. Analysis indicates that the $\mathcal{PT}$-symmetric component of the periodic optical refractive index can modify the grating band structure and hence the effective coupling between the forward and backward waves. Starting from a classical modified massive Thirring model, solitary wave solutions are obtained in closed form. The basic properties of these slow solitary waves and their dependence on their respective $\mathcal{PT}$-symmetric gain/loss profile are then explored via numerical simulations.
\end{abstract}


\maketitle
\section {INTRODUCTION}
Periodic structures play an important role in the general area of optics. As in solid state physics, the periodicity in their refractive index can lead to a succession of photonic band gaps and transmission bands. In many applications, these properties are used to obtain high reflectivities, frequency filtering, and high-dispersion characteristics \cite{b1}. Index gratings, whether in the bulk \cite{b2} or embedded in optical fibers \cite{a3,a4}, are examples of such structures.  Even though in principle these periodic configurations can always be rigorously analyzed using a Floquet-Bloch approach, on many occasions a coupled-mode formalism will suffice. As shown by Kogelnik, this latter formalism is particularly successful when the periodic index perturbation is weak, in which case the coupling between the forward and backward waves occurs over a distance of several wavelengths \cite{a5}. In this regime, the interaction can be described through the so-called slowly varying approximation, which in turn leads to a relatively simple system of coupled equations.\\ 

The behavior of Kerr nonlinear optical periodic systems was first addressed in 1979 under continuous-wave conditions in conjunction with optical bistability \cite{a6}. Few years later, it was realized that this same system can also support a special class of soliton solutions-the so called Bragg solitons \cite{a611,a7,a8}. Unlike optical solitons propagating in nonlinear dispersive fibers, this family of waves is made possible by nonlinearly interlocking both the forward and backward propagating modes \cite{a9,*a92,a911}. In doing so, these wavepackets "open up" a defect band within the forbidden band gap thus allowing energy transport. Given that under Bragg conditions this propagation is linearly forbidden (the grating acts like a distributed mirror), the resulting propagation can be very slow as in the case of self-induced transparency. In general, the velocity of Bragg solitons can range from zero (fully immobile light) to $c/n$, depending on excitation conditions. We would like to emphasize that so far, this class of waves has been primarily investigated in conservative systems. The question arises, as to how they will behave in non-conservative environments, especially in the presence of linear gain or loss.\\ 

Parity-time ($\mathcal{PT}$) symmetry in optics has recently attracted considerable attention \cite{a10,a11,a12,a13,a131,a14,a15,a16,a17,*a172,a18,a19,a20,a21,a22,a23,a231,a232,a2321}. While $\mathcal{PT}$ symmetry was first explored within the quantum domain \cite{a24,a25,a26,a27}, it is in optics that has found a straightforward realization where its implications can be directly observed and studied \cite{a12,a13}. As shown in \cite{a10,a11}, an optical system obeys $\mathcal{PT}$ symmetry provided that its complex refractive index distribution $n(\bm{r})=n_R (\bm{r})+in_I (\bm{r})$ satisfies the condition $n^* (\bm{r})=n(-\bm{r})$ . In other words, the real index profile must be an even function of position while the gain/loss must be odd. It can be shown that for such structures, a real propagation constant (eigenenergies in the Hamiltonian language) exists for some range ({\it exact $\mathcal{PT}$-symmetric phase}) of the gain/loss coefficient. For larger values of this coefficient the system undergoes a {\it spontaneous symmetry breaking}, corresponding to a transition from real to complex spectra ({\it broken $\mathcal{PT}$-symmetric phase}). The phase transition point, shows all the characteristics of an {\it exceptional point} (EP) singularity. Abrupt $\mathcal{PT}$-symmetry breaking has been recently observed in both active and passive experimental arrangements \cite{a12,a13,a232}. In addition non-reciprocity in propagation as well as double refraction and energy oscillations have been predicted in periodic lattices and coupled structures. The possibility of unidirectional invisibility was put forward in linear and nonlinear $\mathcal{PT}$-symmetric gratings and the 
properties of $\mathcal{PT}$-symmetric scatterers and lasers were also discussed \cite{a20,a21}. Finally, along somewhat different lines, the prospect for static optical solitons in $\mathcal{PT}$-periodic arrays has been considered in several studies \cite{a28,a29,a30,a31,a32}.\\

In this work we demonstrate that a new family of optical Bragg solitons is possible in Kerr nonlinear $\mathcal{PT}$-symmetric periodic structures. Starting from a classical modified massive Thirring model \cite{a33}, solitary wave solutions are obtained in closed form. The basic properties of these slow solitary waves and their dependence on their respective $\mathcal{PT}$-symmetric gain/loss profile are explored and pertinent numerical simulations are carried out to elucidate their behavior. We also show that at the exceptional point, the evolution equations decouple, thus allowing a special class of solutions.\\

\section{THEORETICAL ANALYSIS} 
\begin{figure}[h!]
\begin{center}
\includegraphics[width=3.3in]{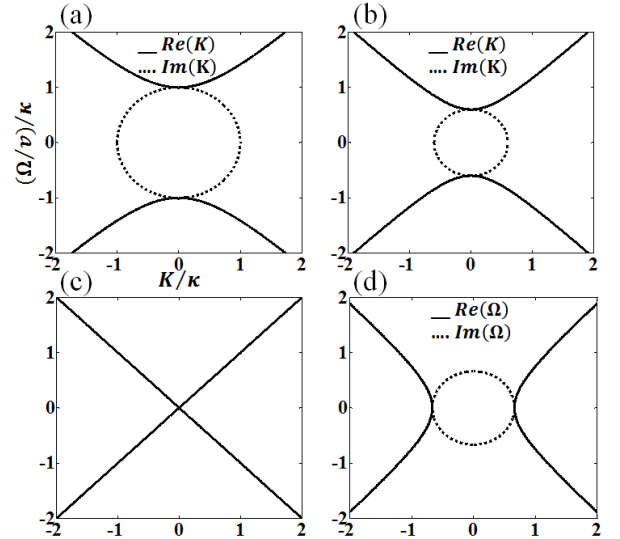}
\caption{(Color online) Band structure of a $\mathcal{PT}$-symmetric periodic grating (linear case) for different ratios of $g/\kappa$; (a) $0$, (b) $0.8$, (c) $1$, (d) $1.2$}
\label{fig1}
\end{center}
\end{figure}

\tp{We} begin our work by considering a $\mathcal{PT}$-symmetric optical grating having a periodic complex refractive index distribution: let us consider a fiber with the following refractive index of the core:
\begin{equation}
\label{eq1}
n=n_0+n_{1R} cos(\frac{2 \pi}{\Lambda}z) +in_{1I} sin(\frac{2 \pi}{\Lambda}z)+n_2 |E|^2   
\end{equation}
In this profile the first term stands for the refractive index of the background material involved while the three other terms are considered to be small perturbations on $n_0$; the second term describes periodic Bragg grating, the third term represents the superimposed complex $\mathcal{PT}$ potential (gain or loss) and the last term accounts for the Kerr nonlinearity. We now express the solution as a sum of forward and backward propagating waves:
\begin{equation}
\label{eq2}
E=E_f (z,t)  exp[i(\beta_0 z-\omega_0 t)]+E_b (z,t)  exp[-i(\beta_0 z+\omega_0 t)]
\end{equation}
where $\omega_0=2\pi c/\lambda_0$ is the carrier angular frequency, $\lambda_0$ is the free space wavelength and $\beta_0=n_0 \omega_0/c$ is the unperturbed propagation constant. Finally $E_f (z,t)$ and $E_b (z,t)$ represent slowly varying amplitudes for the forward and backward waves respectively. In this case, it can be directly shown that the two slowly varying envelope functions satisfy the following coupled wave equations:
\begin{subequations}
\label{eq3} 
\begin{center}
\begin{eqnarray}
\begin{split}
+i\left (\frac{\partial E_f}{\partial z}+\frac{1}{v}  \frac{\partial E_f}{\partial t}\right )&+\left (\kappa+g\right ) e^{-i2\delta z} E_b\\&+\gamma (|E_f |^2+2|E_b |^2 ) E_f=0,\label{eq3a}
\end{split}
\\
\begin{split}
-i\left (\frac{\partial E_b}{\partial z}-\frac{1}{v}  \frac{\partial E_b}{\partial t}\right )&+\left (\kappa-g\right ) e^{+i2\delta z} E_f\\&+\gamma (|E_b |^2+2|E_f |^2 ) E_b=0,\label{eq3b}
\end{split}
\end{eqnarray}
\end{center}
\end{subequations}
In the above equations $v=c/n_0$ is the wave velocity in the background material, $\kappa=\pi n_{1R}/\lambda_0$ is the coupling coefficient arising from the real Bragg grating itself, and $g=\pi n_{1I}/\lambda_0$ is the anti-symmetric coupling coefficient arising from complex $\mathcal{PT}$ potential term. In addition, $\delta=(n_0/c)(\omega_0-\omega_B)$ is a measure of detuning from the Bragg angular frequency $\omega_B=\pi c/(n_0 \Lambda)$ and $\gamma=n_2 \omega_0/c$ is the self-phase modulation constant.\\ 
In the linear regime, the properties of Eq.~(\ref{eq3}) can be readily understood by using the following gauge transformation, $E_f=Fe^{-i\delta z} e^{iv\delta t}$, $E_b=Be^{i\delta z} e^{iv\delta t}$ , in which case one obtains:
\begin{subequations}
\label{eq4} 
\begin{eqnarray}
&+i\left (\frac{\partial F}{\partial z}+\frac{1}{v}  \frac{\partial F}{\partial t}\right )+\left (\kappa+g\right )B=0,\label{eq4a}
\\
&-i\left (\frac{\partial B}{\partial z}-\frac{1}{v}  \frac{\partial B}{\partial t}\right )+\left (\kappa-g\right )F=0,\label{eq4b}
\end{eqnarray}
\end{subequations}
By assuming time harmonic solutions of the form, $(F,B)=(F_0,B_0 )  exp\left ( i(Kz-\Omega t)\right ) $ we arrive at the dispersion relation:
\begin{equation}
\label{eq5}
K^2=\frac {\Omega^2}{v^2} -(\kappa^2-g^2 )
\end{equation}
The effect of the $\mathcal{PT}$-symmetric term arising from $g$ on the overall dispersion characteristics of this Bragg grating is obvious. In essence, its presence can effectively shift the photonic band gap as illustrated in Fig.~\ref{fig1}, for different ratios of $g/\kappa$.
In Fig. ~\ref{fig1}, the dispersion properties of this periodic $\mathcal{PT}$ grating are depicted for three different regimes, depending on the ratio of $g/\kappa$; b) for $g<\kappa$ (below $\mathcal{PT}$-symmetry breaking threshold) the band structure has essentially the shape of an ordinary Bragg grating-with the photonic band gap reduced, c) for $g=\kappa$ (at the $\mathcal{PT}$ threshold or exceptional point) the band gap is closed and the dispersion curve is identical to that expected from the homogeneous background material, and d) for $g>\kappa$ (above threshold) where no band gap exists and the dispersion relation is totally different in shape. As Fig. 1(d) illustrates, above the $\mathcal{PT}$-symmetry breaking threshold, around the origin, there is always a range of wavevectors associated with complex frequencies. As we will see, this latter observation explains why in this case field configurations can grow/decay exponentially with propagation distance. In addition, in this same regime the group velocity is always larger than velocity of light within the background material.
In this work, we mainly restrict our attention in the first range, i.e., we will assume  that the $\mathcal{PT}$ grating will be operated below the $\mathcal{PT}$ threshold where the entire frequency spectrum is real.
\section{NONLINEAR DYNAMICS AND SOLITARY WAVE SOLUTIONS}
In this section we investigate the existence of solitary wave solutions for the coupled wave Eqs.~\ref{eq3}. To do so, we exploit the existing similarity between Eqs.~\ref{eq3} and of that of the massive Thirring model \cite{a33}. By introducing the two parameters $\rho=\sqrt{(\kappa-g)/(\kappa+g)}$ and $\kappa_\rho=\sqrt{\kappa^2-g^2}$ and by employing the gauge transformations $E_f=Fe^{-i\delta z} e^{iv\delta t}$, $E_b=\rho Be^{i\delta z} e^{iv\delta t}$, these coupled wave equations can be written in the following form:
\small
\begin{subequations}
\label{eq6} 
\begin{eqnarray}
&+i\left (\frac{\partial F}{\partial z}+\frac{1}{v}  \frac{\partial F}{\partial t}\right )+\kappa_\rho B+\gamma \left (|F|^2+2\rho^2|B|^2 \right ) F=0,\label{eq6a}
\\
&-i\left (\frac{\partial B}{\partial z}-\frac{1}{v}  \frac{\partial B}{\partial t}\right )+\kappa_\rho F+\gamma \left (\rho^2|B|^2+2|F|^2 \right ) B=0. \label{eq6b}
\end{eqnarray}
\end{subequations}
\normalsize
We note that the above mentioned gauge transformation is only valid when $\kappa>g$, e.g. below the $\mathcal{PT}$ threshold point. As a next step we consider a solution of the form:
\begin{equation}
\label{eq7}
(F,B)=\alpha(\psi_f,\psi_b ) e^{i\eta (z,t)}
\end{equation}
where the constant $\alpha$ and the function $\eta(z,t)$ remain to be determined. On the other hand, $(\psi_f,\psi_b)$ represent solutions to the Thirring model \cite{a7,a8,a9,*a92}:
\begin{subequations}
\label{eq8} 
\begin{eqnarray}
\psi_f=&+\sqrt{\frac{\kappa_\rho}{2\gamma}}  \frac{1}{\Delta}  sin(\sigma) e^{i\Phi}  sech\left (\theta-i\frac{\sigma}{2} \right ),\label{eq8a}
\\
\psi_b=&-\sqrt{\frac{\kappa_\rho}{2\gamma}} \Delta  sin(\sigma) e^{i\Phi}  sech\left (\theta+i\frac{\sigma}{2} \right ),\label{eq8b}
\end{eqnarray}
\end{subequations}
where $\Phi$ and $\theta$ are functions of $z$ and $t$ defined as follows:
\begin{equation}
\label{eq9}
\theta=\kappa_\rho sin(\sigma) \frac{z-vmt}{\sqrt{1-m^2}}
\end{equation}
\begin{equation}
\label{eq10}
\Phi=\kappa_\rho cos(\sigma) \frac{mz-vt}{\sqrt{1-m^2}}
\end{equation}
In the above, the dimensionless quantity $m$ is defined as $m=(1-\Delta^4)/(1+\Delta^4)$ and finally $\Delta$ and $\sigma$ ($0<\sigma<\pi$) are free parameters. After inserting these solutions into Eq.~\ref{eq6} we then obtain:
\small
\begin{subequations}
\label{eq11} 
\begin{eqnarray}
\frac{d\eta}{d\theta}=&+\left (\frac{1}{2}\frac{\alpha^2}{\Delta^4}+\rho^2\alpha^2-1\right ) sin(\sigma) \left |sech(\theta-i\frac{\sigma}{2})\right |^2,\label{eq11a}
\\
\frac{d\eta}{d\theta}=&-\left (\frac{1}{2}\alpha^2\rho^2\Delta^4+\alpha^2-1\right ) sin(\sigma) \left |sech(\theta-i\frac{\sigma}{2})\right |^2 \label{eq11b}
\end{eqnarray}
\end{subequations}
\normalsize
A valid solution of Eqs.~\ref{eq11} requires that both sides are equal. This condition in turn determines the unknown coefficient $\alpha$:
\begin{equation}
\label{eq12}
\alpha=\left (\frac{1+\rho^2}{2}+\frac{1+\rho^2\Delta^8}{4\Delta^4} \right )^{-1/2}
\end{equation}
Finally $\eta$ can then be obtained by integrating either one of Eqs.~\ref{eq11}:
\begin{equation}
\label{eq13}
\eta(\theta)=2\left (\frac{\alpha^2}{2\Delta^4}+\rho^2\alpha^2-1\right )tan^{-1}\left (tanh(\theta)tan\left (\frac{\sigma}{2} \right ) \right )
\end{equation}
Here it is worth discussing the velocity and instantaneous frequency associated with this soliton solution. According to Eqs.~(\ref{eq8}) and (\ref{eq9}) the soliton velocity can be readily obtained from:
\begin{equation}
\label{eq131}
v_s=\frac{1-\Delta ^4}{1+\Delta ^4} v
\end{equation}
Hence the soliton velocity can reach any value  between zero ($\Delta =1$) and the group velocity in the background medium ($\Delta =0$). Using an amplitude and phase representation of Eqs.~(\ref{eq7}) and (\ref{eq8}), the corresponding phase of this soliton solutions could be written as,
\begin{equation}
\label{eq132}
\Xi =\eta + \Phi \pm {tan}^{-1}\left (tanh(\theta)tan\left (\frac{\sigma}{2}\right )\right )
\end{equation}
where the plus and minus signs correspond to the forward $F$ and backward component $B$ respectively. Note that these phases are obtained after the aforementioned gauge transformation. Hence to obtain the actual phases for the forward and backward waves ($E_f$,$E_b$) the term  $v\delta t\mp \delta z$ must be added to these phases respectively. The instantaneous angular frequency can then be obtained from a first order term Taylor series expansion of the respective phase of Eq.~(\ref{eq132}):
\scriptsize
\begin{equation}
\label{eq133}
\begin{split}
\Omega_s&=\frac{{\kappa}_{\rho}v}{\sqrt{1-m^2}}cos(\sigma)\\
&+\frac{{\kappa}_{\rho}v}{\sqrt{1-m^2}}sin(\sigma)\\
&\times\left (\frac{\alpha^2}{2\Delta^4}+\rho^2\alpha^2-1\pm 0.5\right )\frac{2m~tan\left (\frac{\sigma}{2}\right )sech^2\left (\theta(z,t=0)\right )}{1+tan^2\left (\frac{\sigma}{2}\right )tanh^2\left (\theta(z,t=0)\right )} 
\end{split}
\end{equation}
\normalsize
Given that a gauge transformation was used, the quantity $v\delta$ must be subtracted from the result of Eq.~(\ref{eq133}), which is measured with respect to the carrier frequency. Thus the total instantaneous angular frequency of this soliton solution is given by $\omega_s=\Omega_s-v\delta+\omega_0=\Omega_s+\omega_B$. According to the linear dispersion analysis used in the previous section, the frequency band gap for the $\mathcal{PT}$-symmetric grating can be obtained from $-\kappa_\rho v<\Omega<\kappa_\rho v$. Therefore, based on Eq.~(\ref{eq133}) the soliton frequency $\Omega_s$ may or may not lie in the band gap.\\
Up to this point, the solutions were obtained for $\kappa>g$, i.e. before the $\mathcal{PT}$ symmetry is broken. On the other hand, at exactly the $\mathcal{PT}$-symmetry breaking point ($\kappa=g$), the effective coupling coefficient $\kappa_\rho$ goes to zero. In this case, the evolution equations are not completely decoupled and can be more effectively treated in the original set of variables. By introducing the gauge transformations $E_f=Fe^{-i\delta z} e^{iv\delta t}$, $E_b=Be^{i\delta z} e^{iv\delta t}$ , the coupled wave Eqs.~\ref {eq3} reduce to:
\small
\begin{subequations}
\label{eq14} 
\begin{eqnarray}
&+i\left (\frac{\partial F}{\partial z}+\frac{1}{v}  \frac{\partial F}{\partial t}\right )+2\kappa B+\gamma \left (|F|^2+2|B|^2 \right ) F=0,\label{eq14a}
\\
&-i\left (\frac{\partial B}{\partial z}-\frac{1}{v}  \frac{\partial B}{\partial t}\right )+\gamma \left (|B|^2+2|F|^2 \right ) B=0,\label{eq14b}
\end{eqnarray}
\end{subequations}
\normalsize
The linear coupling term between the forward and backward waves now breaks the symmetry in the evolution equations. Note that there is no energy transfer from the forward wave to the backward but the backward wave facilitates energy transfer to the forward. This can be better understood by considering the general solution of Eq.~\ref {eq14b}, given by:
\begin{equation}
\label{eq15}
B=b(y)exp\left (-i\gamma\left [ b^2(y)x+2\int_{0}^{x}|F|^2d\xi\right ] \right )
\end{equation} 
where $x=\frac{1}{2}(z-vt)$, $y=\frac{1}{2}(z+vt)$ are forward and backward propagation coordinates and b is an arbitrary function.  On the other hand Eqs.~\ref{eq14}  admit a trivial solution when $B=0$. In this latter case, Eq.~\ref{eq14a} reduces to that describing a forward propagating wave in the presence of nonlinear self-phase modulation, which admits the following solution:
\begin{equation}
\label{eq16}
F=f(x)exp\left (i\gamma f^2(x)y \right )
\end{equation} 
where $f$ is an arbitrary function. In the other words, in this regime the intensity profile of the forward propagating wave remains invariant during propagation while no energy is transferred to the backward mode. 
\section{NUMERICAL RESULTS}
In this section we exemplify our results through numerical simulations of Eqs.~(\ref{eq6}). The numerical methods used for solving the coupled wave equations presented are based on finite difference methods using different discretizing approaches in order to account for numerical stability \cite{a8,a9,*a92}. Here for discretization we use Euler's method that is based on a first order approximation for both temporal and spatial derivatives. In this case stability would not be an issue as long as the temporal step size is way smaller than the spatial step size.  
\begin{figure} [h!]
\begin{center}
\includegraphics[width=3.5in]{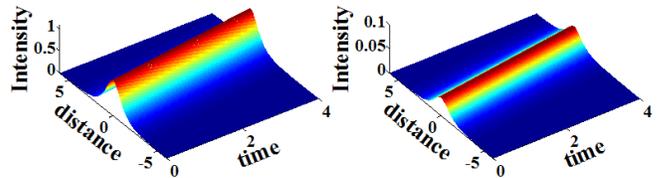}
\caption{(Color online) Propagation dynamics of a solitary wave solution in a $\mathcal{PT}$-symmetric Bragg structure; intensity evolution for both the forward (left) and backward waves (right) during propagation.}
\label{fig2}
\end{center}
\end{figure}
First we investigate the behavior of the solitary wave solution given by Eqs.~(\ref{eq7}-\ref{eq13}). Figure \ref{fig2} depicts the corresponding propagation dynamics of this solution for both the forward and backward waves. According to this figure, these two components propagate at a common velocity and they have the same profile (except from a scaling factor that is clear from Eqs.~(\ref{eq8})). In this numerical example $g/\kappa=0.8$, and the space-time coordinates are normalized as follows: $Z=\kappa z$ and $T=\kappa vt$. In addition the forward and backward electric fields are also here normalized with respect the quantity $E_0=\sqrt{\kappa/\gamma}$. The parameter $\sigma$ that determines the beam width of these solitons is taken to be $\pi/2$, and parameter $\Delta$ that determines the common velocity of the two constituent waves is taken to be $0.8$.
\begin{figure}[t,b,h]
\includegraphics[width=3.5in]{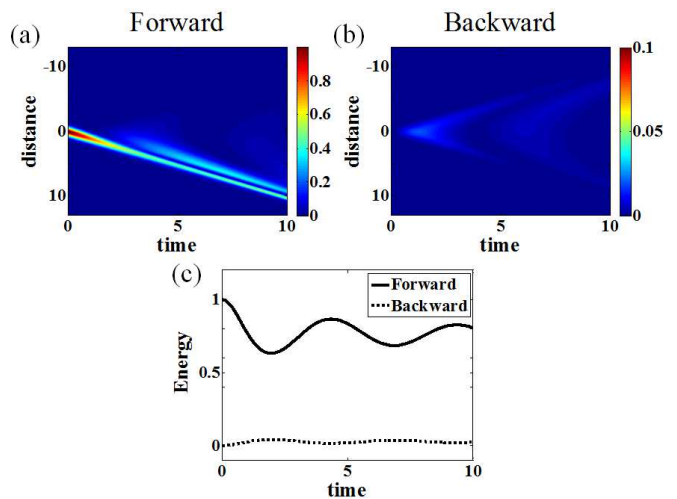}
\caption{(Color online) Propagation dynamics of a Gaussian wavepacket when injected only in the forward direction when the $\mathcal{PT}$ grating is operated below the $\mathcal{PT}$-symmetry breaking threshold. (a),(b) depict the forward and backward components respectively and (c) the associated energy as a function of normalized time.}
\label{fig3}
\end{figure}
\begin{figure}[t,b,h]
\includegraphics[width=3.5in]{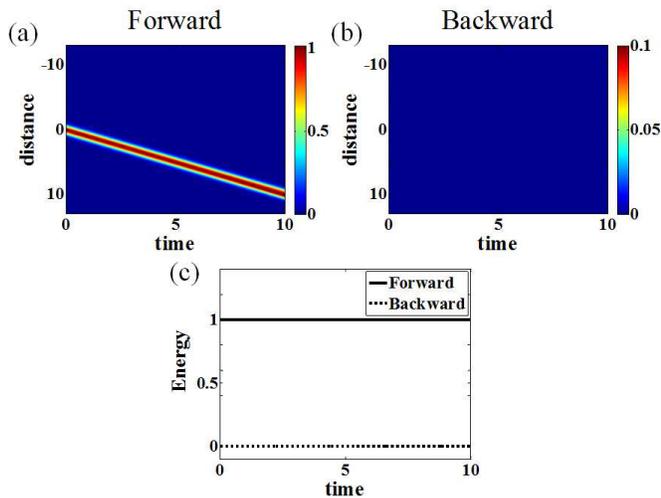}
\caption{(Color online) The same as figure \ref{fig3} when the $\mathcal{PT}$ grating is operated at the $\mathcal{PT}$-symmetry breaking threshold. (a),(b) depict the forward and backward components respectively and (c) the associated energy as a function of normalized time.}
\label{fig4}
\end{figure}
Figures (\ref{fig3},\ref{fig4}) on the other hand show the evolution of a Gaussian pulse when it excites only the forward wave within such a $\mathcal{PT}$-symmetric Bragg grating, for two different cases: below the $\mathcal{PT}$-symmetry breaking point and at threshold. In these simulations $g/\kappa$ is set to be $0.8$, $1$ respectively. In these figures the total energy of each component that is proportional to $\int_{-\infty}^{\infty}|H(z,t)|^2dz$ (where $H$ is either a forward or a backward wave) is also plotted as a function of time. In the case of $\mathcal{PT}$-symmetric soliton solutions this quantity is constant with propagation.\\
According to Fig. \ref{fig3}, below $\mathcal{PT}$ threshold there is an oscillatory power exchange between the forward and backward waves. In this same regime, by increasing the amplitude of the imaginary potential (amplitude of gain or loss), the rate of this energy exchange decreases.  Figure \ref{fig4} on the other hand, shows that the forward Gaussian wave remains unchanged during propagation while the backward wave is not excited at all. This is in agreement with our previous discussion, as expected from equation (\ref{eq16}). This is because there is  no energy coupling between the forward and backward wave.\\
\section{CONCLUSIONS}
In this work we have demonstrated that a new family of optical Bragg solitons is possible in Kerr nonlinear $\mathcal{PT}$-symmetric periodic structures. 
By considering the connection to the classical modified massive Thirring model, solitary wave solutions were obtained in closed form. The 
basic properties of these slow solitary waves and their dependence on their respective $\mathcal{PT}$-symmetric gain/loss profile were explored and 
pertinent numerical simulations were carried out to elucidate their behavior. Finally, of interest will be to examine if similar concepts can be applied in other periodic
structures as for example in nonlinear optical mesh lattices \cite{a40,a41}.\\
\begin{acknowledgments}
This research was supported by a AFOSR No. FA 9550-10-1-0561 grant, by an AFOSR No. FA 9550-10-1-0433 grant, and by an NSF ECCS-1128571 grant.
V.K.'s work was supported via AFOSR LRIR 09RY04COR and via the OSD Metamaterials Insert.
\end{acknowledgments}
\bibliography{GF}
\end{document}